# Superconducting Order Parameters in overdoped BaFe$_{1.86}$Ni$_{0.14}$As$_2$ Revealed by Multiple Andreev Reflection Spectroscopy of Planar Break-Junctions

T. E. Kuzmicheva[*], S. A. Kuzmichev[**,*], K.S. Pervakov[*], V.A. Vlasenko[*]

[*]Lebedev Physical Institute, Russian Academy of Sciences, 119991 Moscow, Russia

[**]Lomonosov Moscow State University, Faculty of Physics, 119991 Moscow, Russia

e-mail:kuzmichevate@lebedev.ru

Using multiple Andreev reflection (MAR) spectroscopy, we studied superconductor - normal metal - superconductor (SnS) semiballistic contacts with incoherent transport in overdoped pnictide superconductors BaFe$_{1.86}$Ni$_{0.14}$As$_2$ with critical temperature $T_c \approx 12$ K. We directly determined an anisotropic large gap with the characteristic ratios $2\Delta_L^{out}(0)/k_B T_c \approx 6$ and $2\Delta_L^{in}(0)/k_B T_c \approx 4$, almost isotropic small gap with $2\Delta_S(0)/k_B T_c \approx 2$, and the temperature dependence of the resolved superconducting order parameters. Above the large gap position $2\Delta_L$, a fine dI(V)/dV structure fully vanishing at $T_c$ and possibly caused by a resonant coupling with a bosonic mode was observed.

## Introduction

Ba(Fe,Ni)$_2$As$_2$ pnictides belong to a so called 122 structural type of the iron-based superconductors [1]. Its layered structure consists of quasi-two-dimensional superconducting Fe-As blocks alternating by Ba planes along the c-direction. At low temperatures, the parent BaFe$_2$As$_2$ compound shows spin density wave (SDW) ground state, covered by nematic ordering phase that develops until temperatures about 140 K [2,3]. Under electron, hole, or isovalent substitution, the SDW ordering is gradually suppressed, in advance with an emergence of superconducting dome. For the nickel-substituted BaFe$_{2-x}$Ni$_x$As$_2$ compound, the optimal x=0.1 concentration gains the critical temperature up to $T_c \approx 20$ K. A nematic fluctuations region that envelops the nematic phase and further expands even to overdoped superconducting phase, was reported in the works [3-6]. Despite intensive studies of the 122 family pnictides, an interplay between magnetic, orbital, and nematic orderings, and their influence to superconductivity are still highly debatable [6-12].

Angle-resolved photoemission spectroscopy (ARPES) experiment [13-16] shows that the Fermi level $E_F$ is crossed by several bands close to a Lifshitz transition. In electron doped pnictides of the 122 family, the Fermi surface contains quite nested hole barrels near the Γ point and electron barrels near the M point of the folded Brillouin zone. Below $T_c$, as generally recognized, two superconducting condensates are developed with the large gap $\Delta_L$ and the small gap $\Delta_S$ order parameters.

To date, the structure of the superconducting order parameter in Ni-substituted Ba-122 pnictides is poorly studied in the experiment. A substantial gap anisotropy in overdoped compounds was claimed in [17]. For optimally doped crystals, bulk probes of specific heat and London penetration depth [18-20] showed the best fit of the experimental data with two isotropic gaps with the characteristic ratios $2\Delta_L(0)/k_BT_c$ = 3.6-3.8 and $2\Delta_S(0)/k_BT_c$ = 1.3-1.9. Infrared reflection Fourier spectroscopic studies of $BaFe_{1.9}Ni_{0.1}As_2$ films reported the characteristic ratios up to 4.3 and 2 for the large and the small gap, respectively [21]. In point contact Andreev reflection (PCAR) probes [22], the ratios as high as $2\Delta_L(0)/k_BT_c \approx 12$ and $2\Delta_S(0)/k_BT_c \approx 4.6$ were obtained, whereas the overgap dynamic conductance features were attributed to a bosonic resonance. Unfortunately, the large broadening parameter $\Gamma$, comparable to the gap value (in particular, in [22], $\Gamma \approx \Delta_0/3$ was estimated, and even $\Gamma(\omega > \Delta_0^h)$ = (1-3)$\omega$ scattering rate was obtained for the inner hole pocket in works by Fink et al. [23,24]), seems a typical feature of the iron-based superconductors, thus making the extracted energy parameters ambiguous. Our earlier studies of $BaFe_{1.9}Ni_{0.1}As_2$ single crystals [19,25,26] resolved a nodeless large gap with a substantial 25\%-35\% anisotropy and isotropic small gap with the characteristic ratios $2\Delta_L(0)/k_BT_c \approx 4$-6 and $2\Delta_S(0)/k_BT_c \approx 2$, respectively. The directly obtained temperature dependences of both order parameters showed a moderate interband coupling, whereas the $\Delta_L$ anisotropy remained almost constant until $T_c$.

The pairing glue in iron-based superconductors is still debatable. In order to describe their multiple-band superconductivity, several models were suggested. The $s^{\pm}$-model of spin-fluctuation-mediated repulsion [8] predicts a sign-reversed superconducting order parameter, whereas an $s^{++}$ gap develops when considering the coupling through orbital fluctuations enhanced by phonons [27,28], or a strong intraband coupling via phonons combined with a spin-fluctuation mechanism [12], orbital-selective pairing [29,30], or a shape-resonance model [31]. A spin resonance peak at the nesting vector was observed in neutron scattering probes of various iron-based superconductors (for a review, see [9,32,33] and refs. therein). In particular, in $Ba(Fe,Ni)_2As_2$ a linear relation of the spin resonance energy $\varepsilon_0$ with $T_c$ was reported in [33].

Here we present the direct measurements of the superconducting order parameters in overdoped $BaFe_{1.86}Ni_{0.14}As_2$ single crystals with critical temperature $T_c \approx 12$ K using multiple Andreev reflection (MAR) spectroscopy. We determined an anisotropic large gap and isotropic small gap with the characteristic ratios almost similar to those obtained by us earlier in optimally doped $BaFe_{1.9}Ni_{0.1}As_2$ with $T_c \approx 12$ K. Nonetheless, the current data on overdoped samples show some differences in the temperature dependences of the superconducting gaps, as well as possible footprints of a characteristic bosonic mode. Also discussed are the unconventional normal-state properties.

**Experimental details**

The single crystals with the nominal composition $BaFe_{1.86}Ni_{0.14}As_2$ were grown using a ``self-flux'' technique. The synthesis and characterization of the $BaFe_{2-x}Ni_xAs_2$ crystals, as well as their structural, transport, and magnetic probes, are detailed in [25,34-36]. The occurrence of bulk superconductivity at critical temperature $T_c \approx$ 12 K was confirmed by resistive and magnetic susceptibility measurements.

In order to make superconductor - thin normal metal - superconductor (SnS) planar junctions for MAR [37-41] spectroscopy experiment, we used a mechanically-controlled break-junction (MCBJ) technique [42,43]. On the contrary, for the well-known PCAR spectroscopy probes [44], NIS junctions (N is normal metal, I is insulator) described by the Blonder-Tinkham-Klapwijk theory [45] are formed, where a (single) Andreev reflection effect occurs. While MCBJ generally produces symmetric ScS junctions, where c is a constriction separating the banks of one and the same superconductor (with equal superconducting gap), the PCAR technique usually involves the asymmeric contact of a superconductor with a metal (or with another, conventional superconductor having a certain gap), the physical parameters of which (the Fermi velocity, Fermi energy, carrier concentration, etc.) could significantly differ from those of the superconductor under study. This leads to the two following issues. (i) The above mentioned imbalance between the Fermi velocities and Fermi momenta limits the maximum constriction transparency [45]} (the barrier strength Z ≤ 0.3) even in case of absent dielectric layer. (ii) For the ScS junction, temperature does not smear significantly the dI(V)/dV features while $\Delta(T) > k_BT$, as compared to the case of asymmetric junction: contrary to the metallic electrode (or conventional superconducting electrode above its $T_c$), in the both banks of the ScS contact the continuum states are shifted from the Fermi level by $\Delta$.

As well, we emphasize some differences between the MCBJ and PCAR techniques: (a) while PCAR produces the artificial *single* junction, in layered crystals MCBJ can provide both, the single ScS junction and the natural array of equivalent ScS elements (see below); (b) 3-point probe in PCAR experiment versus 4-point probe in MCBJ experiment; (c) during one experiment, PCAR explores the single point on the sample surface, whereas MCBJ facilitates a readjustment of the contact point; (d) PCAR is able to make a junction along both ab- and c-direction, while in the MCBJ experiment only c-oriented junctions are available in layered compounds.

The thin plate-like sample with dimensions about 3 * 1.5 * 0.1 mm$^3$ was attached to a springy sample holder by four-contact pads made of In-Ga paste at room temperature (see Fig. 1 in review [43] for details). After cooling down to T = 4.2 K, the sample holder was gently curved, thus cracking the bulk sample into two halves, with a formation of two cryogenic clefts separated with a weak link, a kind of planar ScS contact (a presumable sketch is shown in the Appendix in the inset to Fig. 10a). The resulting constriction forms far from current and potential contacts, which prevents junction overheating and provides true four-point probe. A layered sample preferably splits along the ab-planes, with an appearance of natural steps and terraces. The height of the step is a multiple of the c unit cell parameter, whereas the terrace size is typically about 10-1000 nm (the corresponding electron microscope image is available as Fig. 2c in [43]). We suppose, the terraces remain flat after the cracking, nonetheless, the curving of the holder could cause some crystal structure defects, in particular, edge dislocations or polysynthetic twinning.

Under fine tuning of the holder curvature, the two cryogenic clefts slide over each other touching onto various terraces, thus minimizing tension effect in the ab-plane, but, however, not excluding some stress along the c-direction. The clefts remain tightly conjuncted during sliding that prevents impurity penetration into the crack and maintains the purity of cryogenic surfaces. Such tuning enables a mechanical readjustment of the constriction (sweeping the area and the resistance of the junction) in order to realize a desired ballistic regime (with the contact dimension d less than the inelastic scattering length l) or diffusive regime (l ≈ d). In kind, such readjustment enables probing up to several tens of junctions in one and the same sample during the same cooldown. The ability of local probing is one of the most advantageous features of the used MCBJ configuration that facilitates collecting a large data statistics in order to check the reproducibility of the superconducting and normal-state parameters. Obviously, only reproducible dI(V)/dV features are caused by bulk effects; on the contrary, defects, surface influence, or geometric resonances (for example, Fiske steps) appear as irreproducible features. Therefore, we select the reproducible data in order to analyze namely the bulk properties.

In the majority of Fe-based superconductors we studied, the constriction is electrically equivalent to a thin layer of normal metal of high transparency (about 80\% - 98 \%), thus providing an observation of multiple Andreev reflections. As a result, the obtained current-voltage characteristics (CVC) and the dI(V)/dV spectra are typical for the high-transparent classical (``long'') SnS-Andreev junction with incoherent transport [37-41,48].

In such ``long'' SnS junction with the corresponding barrier strength Z < 0.5 at temperatures below $T_c$, non-coherent Andreev transport causes an excess current at the whole bias voltage range, which drastically rises at eV → 0 (so called foot area). This behavior differs from that of the so-called ``short'' SNS contact, which transport is coherent (directed by the electron waves interference and Josephson effect, due to the phase difference between both superconducting banks). A total current passing through ``short'' SNS at low bias voltages tends to zero [38,39]. At the same time, the CVC has the supercurrent branch at eV = 0, which is not a case for the incoherent ``long'' SNS junction [40,41,48].

The corresponding dynamic conductance spectrum shows an increase in zero-bias conductance $G_{ZBC}$ (strictly at eV = 0) as compared to the normal one $G_N$ (being the conductance at eV >> 2Δ), and a series of dynamic conductance features called subharmonic gap structure (SGS) [37,40,41,48]. At certain temperature T, the position of SGS dips directly relates to the superconducting gap magnitude Δ(T) as [37,40,41]

eV(T) = 2Δ(T)/n,                                                    (1)

where the subharmonic order n=1, 2,… is the natural number. Unlike probing asymmetric NS and NIS junctions, no fitting of dI(V)/dV is needed in case of SnS contact till $T_c$, which facilitates a precise measurement of temperature dependence of the gap [40,41,48]. With temperature increase, the excess Andreev current and enhanced zero-bias conductance (ZBC) suppresses gradually; with it, the SGS features shift toward zero, and their amplitudes decrease in proportion to the concentration of Cooper pairs [48]. As the contact area turns to the normal state at local critical temperature $T_c^{local}$, all the features caused by MAR transport fully vanish.

In case of multiple-gap superconductor, several series of such features are expected in the dynamic conductance spectrum. A momentum-dependent (extended s-wave) order parameter would cause doublet-like SGS features, whereas the position of two dips forming the doublet corresponds to the maximum and minimum Cooper pair coupling energies [19,43]. In the used MCBJ configuration, the current always flows along the c-direction, thus providing some information about the in-plane anisotropy of the superconducting gap.

For the junctions obtained in layered superconductors with a valuable anisotropy of electrical properties, the condition l > d should be kept along both, ab and c-directions. For our planar MCBJ's, the semiballistic conditions are $l_c > d_c$, and $l^{el}_{ab} > d_{ab}$, where l and $l^{el}$ are the inelastic and elastic scattering lengths, $d_{ab}^2$ is the constriction area to be estimated using Sharvin formula [49]. The out-of-plane inelastic $l_c/d_c$ ratio (both values are taken along the c-direction) roughly corresponds to the number of observed Andreev dips n ≈ $l_c/d_c$ for the case of fully transparent contact and with the presence of inelastic scattering [40,41,48]. For SnS junction with $l_c/d_c$ ≈ 2, expected are the fundamental (n=1) harmonic and a lower-amplitude subharmonic with n=2 [48]. We consider our contacts as ``long'' with the dimensions exceeding the phase coherence length $d_c > 2\xi_0$ since we newer observed neither a current deficiency at eV → 0, nor a supercurrent branch. Also inelastic processes result in carrier phase randomization, which is contrary to the physics of the ``short'' junctions with coherent transport.

Beside single ScS contacts, Andreev arrays with ScSc-…-S structure can be also formed in the MCBJ experiment with layered sample [43,47,50]. The scheme of ScSc-…-S array is shown in the inset to Fig. 10a in the Appendix. Composed of m ScS junctions, such array peers a natural stack of equivalent elements (owning normal and Andreev channels in parallel). Hence, the dI(V)/dV of the array shows the Andreev features at positions being scaled by a natural factor of m as compared to that of single SnS junction:

$$eV_n(T) = m\, 2\Delta(T)/n. \qquad (2)$$

The arrays were typically observed by our group, with constrictions of low-transparency (Josephson regime) [43,51-53] or high-transparency (SnS-Andreev regime) [19,25,26,43,46,47,50], but the nature of the latter is still not understood. Accounting an intrinsic Josephson effect in SISI-…-S arrays observed in high-temperature cuprates and other layered superconductors [51-56], one may suppose an analogous intrinsic MAR effect and a development of Andreev arrays in the steps of the layered crystal structure. Anyway, the position of the features caused by bulk properties would scale with m.

During the mechanical readjustment, one can probe arrays with various area and number of junctions. For each formed array, the number m is *natural* but accidental, so it can be determined by comparing dI(V)/dV curves for various arrays: after scaling the bias voltage axis by m, the dI(I)/dV spectrum turns to that of a single junction. The details of m assignment are clarified in the Appendix, as well as in Fig. 6 in [50]. Using the determined m numbers, each CVC and the corresponding dI(V)/dV spectrum shown in Figs. 1-3 is normalized to the single SnS junction ($V_{norm}$ means V/m, whereas the current axis is kept unnormalized).

At the first glance, it may seem that accidental $m$ inevitably leads to uncertainty of the extracted energy parameters. Contrary to the intuitive expectations, the result is unexpectedly accurate. As shown below in the Appendix (Fig. 10b), for each triple of the gap values measured simultaneously, the dispersion does not exceed ± 3\%. Similar high precision was demonstrated by us earlier in MCBJ probes of $NdO_{0.6}H_{0.36}FeAs$ (see Fig. 6c in [50]). On the other hand, the observed accuracy together with the narrowing of the dI(V)/dV features (see Fig. 6b,d in [50]) indicates the electrical equivalence of the junctions (their $R_N$) constituting the array: obviously, the $R_N$ nonequivalence is less than the obtained ± 3\% gap dispersion.

This value limits an unambiguous m number to be determined as m < 100%/ (2 * 3%) ≈ 16.

The above discussion leads to several assumptions we used in order to interpret the experimental dI(V)/dV spectra:

(a) The normal resistance of the SnS junctions forming each array is almost equal, which could be checked using a dispersion histogram of (simultaneously measured) gap(s) (Fig. 10b, see also Fig. 4).

(b) The physical properties of MCBJ contacts ($R_N$, area, Z, m) as well as the position of the contact point on the cryogenic cleft are random.

(c) For various contacts, only reproducible dI(V)/dV features are caused by bulk effects. Among the large set of data obtained on the samples from the same batch, we select namely reproducible dI(V)/dV characteristics to be analyzed.

(d) As follows from the MAR effect model that accounts inelastic scattering [40], for SnS junction close to diffusive regime with $l_c/d_c$ ≈ 1-3, the fundamental (n=1) harmonic has the largest amplitude [48]. The experimental data presented here, according to our estimates (see below in Section B), corresponds to the junctions being in this regime. The lower-amplitude high-bias features reproducibly observed at |eV| ≈ 8-10 meV are hence interpreted as a fine structure.

(e) Since the superconducting order parameter in the Ba-122 system is generally assumed to be anisotropic, the doublet-like features are expected in the dI(V)/dV spectra.

Summarizing the advantages of MAR spectroscopy of mechanically-controlled break junctions and natural arrays, this technique provides a precise, local and high-resolution probe of the bulk superconducting order parameter, its temperature dependence and any fine structure. In our studies, the dynamic conductance spectra were measured directly by a standard modulation technique [43]. We used a source of direct current mixed with a small-amplitude ac with frequency about 1 kHz. An automatic control changes ac amplitude in order to keep the amplitude of alternating *voltage* from the sample constant, comparing it with the reference ac voltage (by Lock-in selective amplifying of the umbalance signal from the bridge circuit), and gets a feedback from Lock-in nanovoltmeter. The results obtained with this setup are insensitive to the presence of parallel ohmic conduction paths; if any path is present, the dynamic conductance curve shifts along the vertical axis, while the bias stays unchanged.

**Results and discussion**

- **A. Normal-state features**

Figure 1 shows the normal-state current-voltage characteristic and the dynamic conductance spectrum of an MCBJ measured at T = 16 K above $T_c$. The dI(V)/dV spectrum is strongly nonlinear and shows smooth humps (see vertical arrows) resembling a pseudogap widely discussed in cuprates [57] or a charge density wave (CDW) gap proposed even for $Nb_3Sn$ [58], followed by profound deepenings at higher bias voltages. The strong nonlinearity of the I(V) and dI(V)/dV curves above $T_c$ is reproducibly observed for MCBJ's in overdoped $BaFe_{1.86}Ni_{0.14}As_2$, with its gradual flattering at temperatures about 50-70 K. In the superconducting state, the sharp dynamic conductance Andreev dips and enhanced (Andreev) ZBC are presented on such nonlinear background, those fully vanishing at the local $T_c$ and leading to the ZBC dip observation ($G_{ZBC} \approx 0.9\ G_N$).

Since the used MCBJ configuration assures in effective heat sink from the constriction region, and the data shown in Fig. 1 are symmetrical about zero bias, we would not relate such nonlinearity to Schottky barrier or to an overheating effect (that was thoroughly studied in thermal point contacts in $(Ba,Na)Fe_2As_2$ [59]). In our studies, almost similar background with the ZBC dip was reproducibly observed earlier in optimally doped $BaFe_{1.9}i_{0.1}As_2$ single crystals [26]. By contrast, in LiFeAs [60] a pronounced dynamic conductance hump at $T > T_c$ was present at low bias. As for the 1111 family oxypnictides, a minor increase in the dI(V)/dV value toward zero bias above $T_c$ was observed in optimally doped compounds [47,50], whereas the nonlinear dI(V)/dV background with a dip at $eV \to 0$ was typical for underdoped (Sm,Th)OFeAs [47].

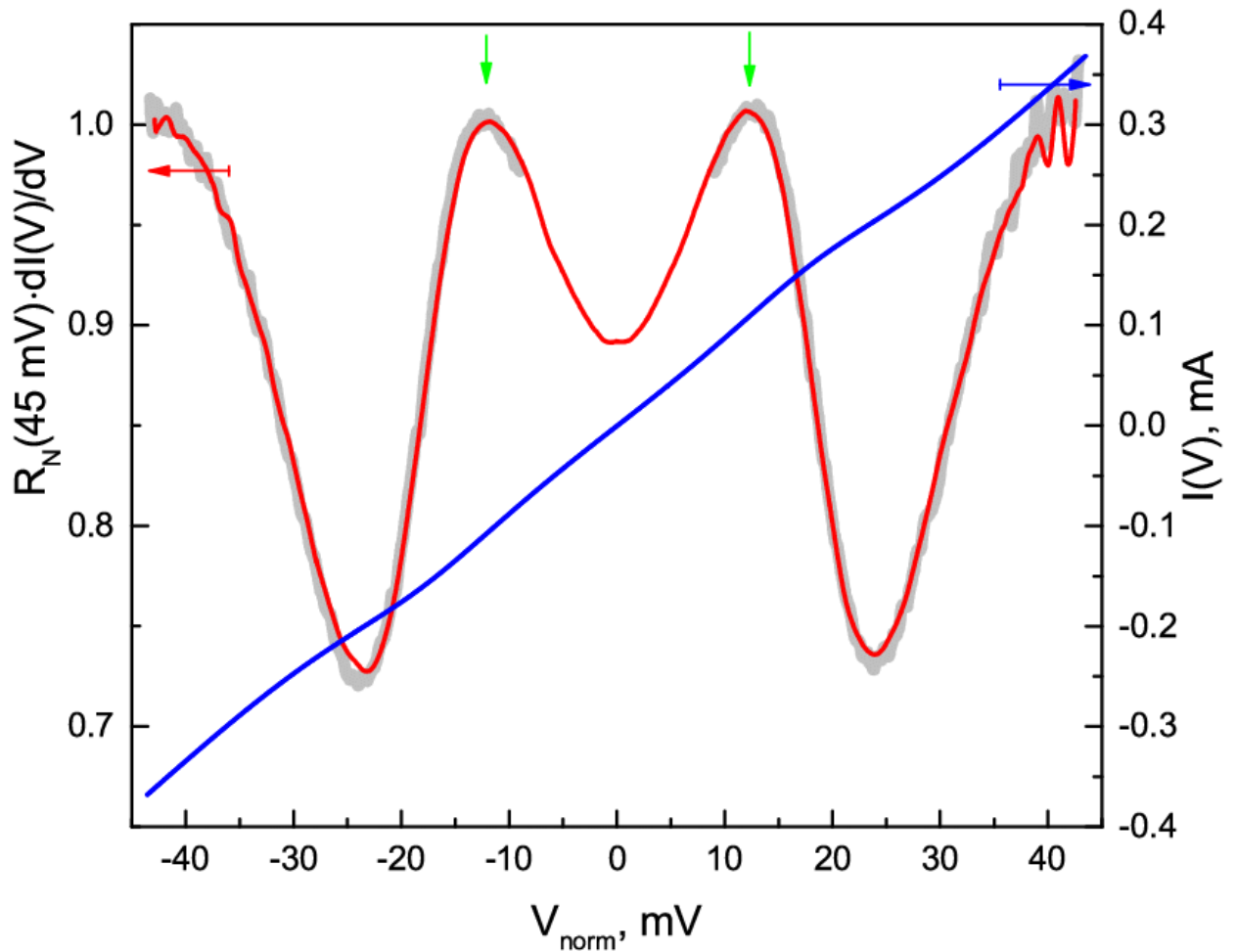

**Fig. 1.** Normal-state electron DOS influence. Current-voltage characteristic (blue line, right axis) and the dynamic conductance spectrum (red curve, left axis) measured at T = 16 K > $T_c$ for NcN array in $BaFe_{1.86}Ni_{0.14}As_2$ single crystal. Vertical arrows point to the humps located at $eV \approx 10$ meV. For comparison, bold gray line shows the fragment (cut at $|eV| = 3\Delta_L(0)$) of the dI(V)/dV for the same junction in the superconducting state measured at T = 4.2 K.

According to the well-known approach suggested by Giaever and Megerle [61], the dynamic conductance of tunnel NcN contact is determined by the energy distribution of the electron density of states (DOS) in the vicinity of the Fermi level $E_F$ (with zero bias corresponding to the Fermi energy). The negative-bias part presents a mirror reflection, thus making the dI(V)/dV of NcN junction always symmetric [61]. For conventional superconductor that satisfies the quasiclassical condition $B_i \sim E_F \gg \Delta_0$, where $B_i$ is the conductive band(s) bottom energy and $\Delta_0$ is the superconducting gap at T=0, the normal DOS near the $E_F$ could be taken constant, hence resulting in a flat $G_N(V)$ and linear (so called ohmic) CVC. In Fig. 1 such normal-state dI(V)/dV features could be caused by nonlinear electron DOS distribution resulting from violated quasiclassical conditions ($B_i \sim 10\, \Delta_0$), that is widely discussed in concern to iron-based superconductors [4,6,13,14,41].

As shown in classical experiments by electron-phonon spectroscopy of ``long'' metallic point contacts, the incoming electrons could be scattered inelastically by phonons in the $l^{inel}$-neighbourhood of the contact. In particular, a backscattering could take place, thus leading to some current deficiency and dI(V)/dV dip at about the relative phonon energy $\hbar\omega_{ph}$. The effect is rather weak but measurable by detecting the features of the $d^2I(V)/dV^2$ [62,63]. As the superconducting gap opens at $T < T_c$, for ScS junction the corresponding energy becomes enlarged by $2\Delta$, as a result, the phonon-caused I(V) and dI(V)/dV features would be shifted by $\hbar\omega_{ph} + 2\Delta_0$ at $T \to 0$ as compared to those measured at $T > T_c$. This is not the case for the data shown Fig. 1: the dI(V)/dV spectrum measured at 4.2 K (bold gray line; the in-gap region was cut for clarity) well coincides the normal-state dI(V)/dV (red line). As well, the observed effect is rather strong, whereas a backscattering that leads to $\approx$ 25% dI(V)/dV decrease is hardly expected. Therefore, the observed normal-state dI(V)/dV nonlinearity could not originate from electron-phonon interaction. Neither could the dip-hump dI(V)/dV features be considered as caused by the spin resonance, since the latter exists in the superconducting state only.

An influence of the nematic fluctuations could be another reason of the appearance of the DOS humps and dips. Recent ARPES studies of $(Ba,K)Fe_2As_2$ crystals revealed a presence of almost ``flat'' replica bands at $\varepsilon \sim$ 10-15 meV $> 2\Delta_0$ below $E_F$ possibly caused by the nematic order [64], those could produce the DOS peaks. In $BaFe_2(As,P)_2$ with isovalent substitution, ARPES data [4] showed the nonequivalent energy position of the bands formed by xz and yz orbitals (thus forming a hump-dip-like DOS) presumably driven by nematic fluctuations; with temperature increase, the lower-energy xz band shifted toward $E_F$, overtaking the static yz band at $T > T_c$. The ``pseudogap''-like DOS peaks above the superconducting gap energies were also predicted in very recent theoretical calculations by Onari and Kontani [65] as resulting from a hidden nematic order above the structural transition. A study of the doping and temperature evolution of the normal-state dI(V)/dV nonlinearity in various iron-based superconductors seems of great importance in concern to an interplay between superconductivity and nematicity, and is an issue of further detailed studies.

- **B. Superconducting Order Parameters at T << $T_c$**

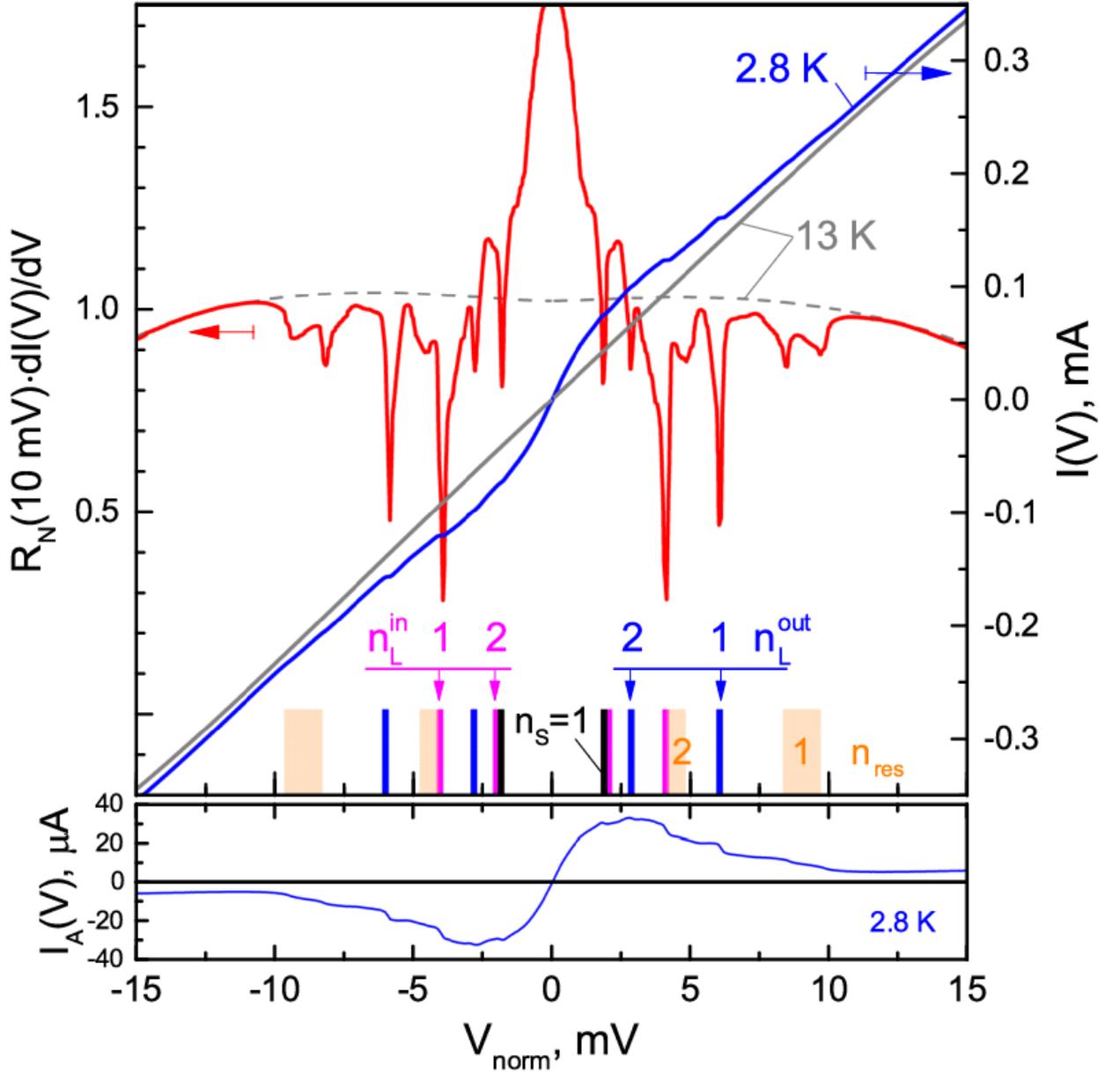

**Fig. 2.** Current-voltage characteristics (right axis) and the dynamic conductance spectrum (red curve, left axis) of SnS-Andreev array (m=11 junctions) measured at T = 2.8 K. The data are normalized by $G_N(\{10 \text{ mV}\} \rightarrow G_N(\infty))$. The vertical dashes show the position of the fundamental (n=1) and the second subharmonics of the large superconducting order parameter $2\Delta_L^{out} \approx 6.1$ meV ($n_L^{out}$ = 1, 2 labels) and $2\Delta_L^{in} \approx 4.2$ meV ($n_L^{in}$ = 1, 2 labels), and of the small gap $2\Delta_S \approx 1.9$ meV ($n_S$ = 1 label). Orange bars show the position of the dips possibly caused by a resonant coupling with a characteristic bosonic mode ($n_{res}$ = 1, 2 labels, see Section D for details). For comparison, the I(V) and dI(V)/dV characteristics measured at T = 13 K > $T_c$ (solid and dashed gray curves) are presented. The lower panel shows an Andreev excess current $I_A(V)$ at 2.8 K versus bias voltage, determined as a difference between the blue and gray CVC's.

As a rough estimation of the in-plane ballistic ratio $l_{ab}^{el}/d_{ab}$, we take the product of bulk resistivity and carrier mean free path $\rho_{ab}l_{ab}^{el} \approx 1.65 * 10^{-3}$ µΩ cm$^2$ estimated for single crystals of sister Ba(Fe,Co)$_2$As$_2$ compound in [66], the resistivity $\rho_{ab} \approx 150$ µΩ cm of the probed BaFe$_{1.86}$Ni$_{0.14}$As$_2$ single crystal just above $T_c$, and R ≈ 42 Ω at eV = 10 meV in the current-voltage characteristic (CVC) shown in Fig. 2 as the normal resistance per one junction for these array. Then we get the in-plane mean free path for the probed sample $l_{ab}^{el} \approx 110$ nm, and estimate the in-plane contact dimension using Sharvin formula [49] with the prefactor for a circular wire (see Eqs. (11,56) in [67]) d = 2a =(8 ρl/(3πR))$^{0.5}$ ≈ 41 nm. The resulting $l_{ab}^{el}/d_{ab}$ ≈ 2.7 along the ab-plane is valid for the realization of the Andreev reflections but is close to the diffusive mode (semiballistic).

The upper panel of figure 2 shows CVC (blue and gray lines, right axis) and dynamic conductance spectrum (red curve, left axis) measured at T = 2.8 K for Andreev array (m=11 junctions) formed in BaFe$_{1.86}$Ni$_{0.14}$As$_2$ single crystal. Note the CVC is (a) symmetric about zero bias, thus indicating the absence of a Schottky barrier, (b) shows no hysteresis that seemingly excludes an appearance of phase-slip centers, and (c) no supercurrent branch at zero bias (measured by using a dc source), which favors an absence of Josephson current through the junction.

As compared to one measured above $T_c$ (gray curve in Fig. 2), the CVC at 2.8 K manifests Andreev excess current in the whole bias range. The voltage dependence of the Andreev current determined as $I_{exc}$(V,2.8 K) = I(V, 2.8 K) - I(V,13 K > $T_c$) is shown in the lower panel. Turning to a constant value at high bias voltages above the gap, $I_{exc}$(V) strongly increases toward eV → 0 (which is contrary to the low-bias current deficiency typical for the ``short'' quantum contacts with phase-dependent current), thus indicating MAR realization in the ``long'' ($d_c > 2\xi_0$) junction with the incoherent transport.

Due to the nonlinear dI(V)/dV background similar to that shown in Fig. 1, it is hard to normalize dynamic conductance to the normal state one $G_N$, as a constant. One of the possible solutions is to take $G_N$(V) at V = $V_{peaks}$ ~ 10 mV (marked by arrows in Fig. 1), since $G_N(V_{peaks}) \rightarrow G_N(\infty)$. Hence, for clarity, the absolute dynamic conductance in Fig. 2 was normalized by that at 10 meV > 2$\Delta_0$.

As compared to the spectrum measured in the normal state just above $T_c$ (gray dI(V)/dV curve), the dynamic conductance at 2.8 K shows a set of minima. The pronounced dips at |eV| ≈ 6.1 meV, and the features at twice lower bias |eV| ≈ 3.05 meV (blue dashes and $n_L^{out}$ = 1,2 label in Fig. 2) could be considered as the fundamental and the second SGS features of the large gap about 3.05 meV. This SGS is overlapped with another intensive dips located at |eV| ≈ 4.2 meV (magenta dashes, $n_L^{in}$=1 label). Although the positions of the two intensive dips roughly satisfy Eq. 1 as n=2 and n=3 subharmonics of the gap value Δ ≈ 6.3 meV, such suggestion is erroneous since there is no fundamental harmonic (which would have larger amplitude) at |eV| ≈ 12.6 meV in the dI(V)/dV-spectrum. We therefore interpret these dips at |eV|=6.1, 4.2 meV as fundamental harmonics of the two energy gap parameters $\Delta_L^{out}$ ≈ 3.05 meV and $\Delta_L^{in}$ ≈ 2.1 meV. The minima observed at positions |eV| ≈ 1.76 meV (black dashes, $n_S$=1 label), as shown below, demonstrate the temperature behaviour differing from that of the large gap subharmonics, thus to be attributed to the distinct, small superconducting gap $\Delta_S$ ≈ 0.88 meV.

The second subharmonic of the $\Delta_L^{in}$ gap expected at $|eV| \approx 2.1$ meV possibly interferes with $n_S=1$ features of the small gap at low temperatures.

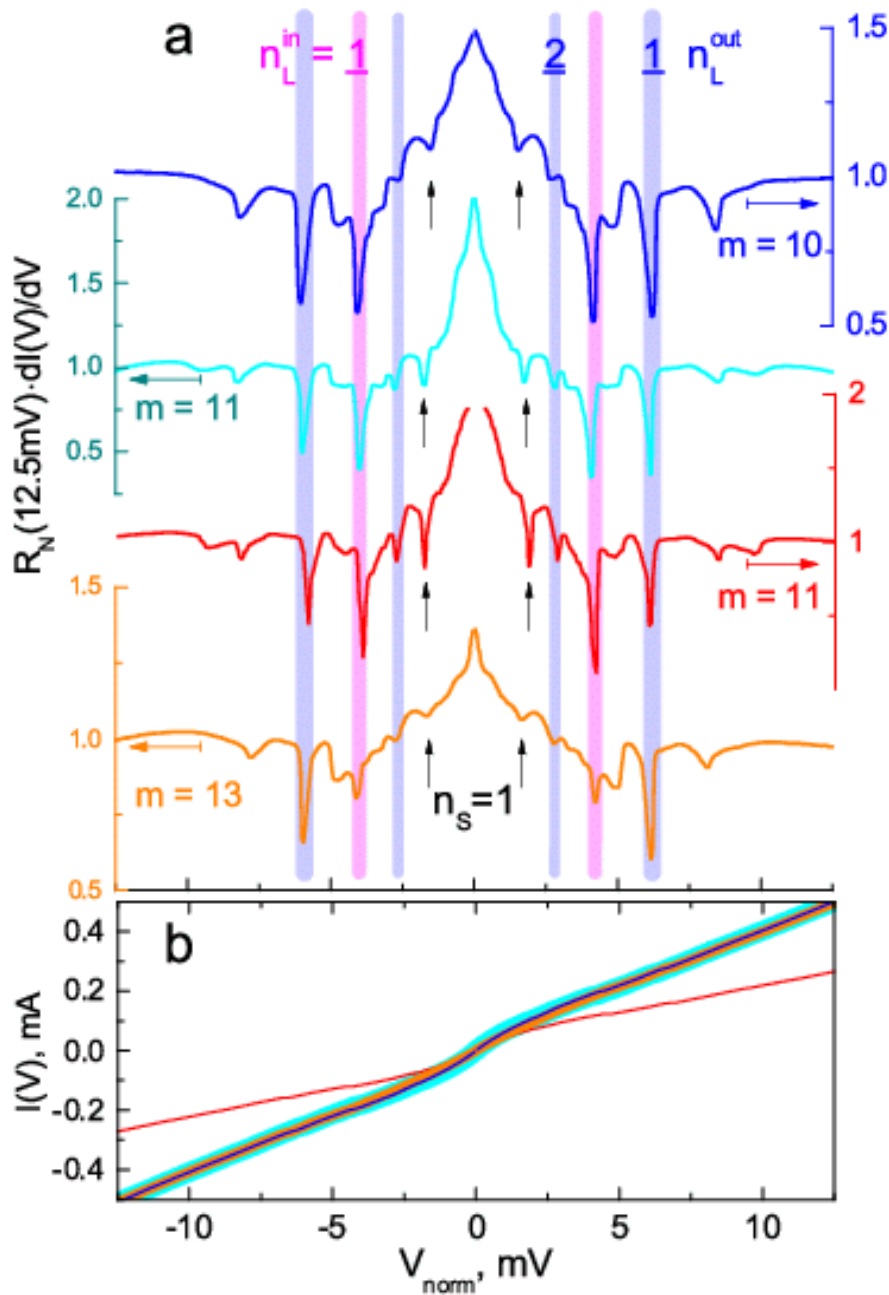

**Fig. 3.** a) Dynamic conductance spectra (upper panels) and (b) current voltage characteristics (shown by the corresponding colors) of SnS-Andreev arrays with various m. The red color curve is the same as in Fig. 2. For each array, the dI(V)/dV is normalized by $G_N(12.5\text{ mV}) \gg 2\Delta_i$, and the bias voltage is normalized by m as $V_{norm} = V/m$. Vertical bars point to the position of the main features caused by the superconducting order parameters (the same notification as in Fig. 2).

Accounting the gap anisotropy predicted by the theoretical models [6,8,27,29,30] and widely detected in various experiments with the 122 family pnictides, the observed two characteristic energy parameters $\Delta_L^{in}$ and $\Delta_L^{out}$ could be interpreted as the edges of the momentum-dependent large superconducting gap. On the other hand, one should not exclude a developing of two isotropic superconducting gaps $\Delta_L^{in}$ and $\Delta_L^{out}$ at different Fermi surface sheets. In order to distinguish unambiguously between these two cases, one needs in extension of the available MAR models with the momentum-dependent gap.

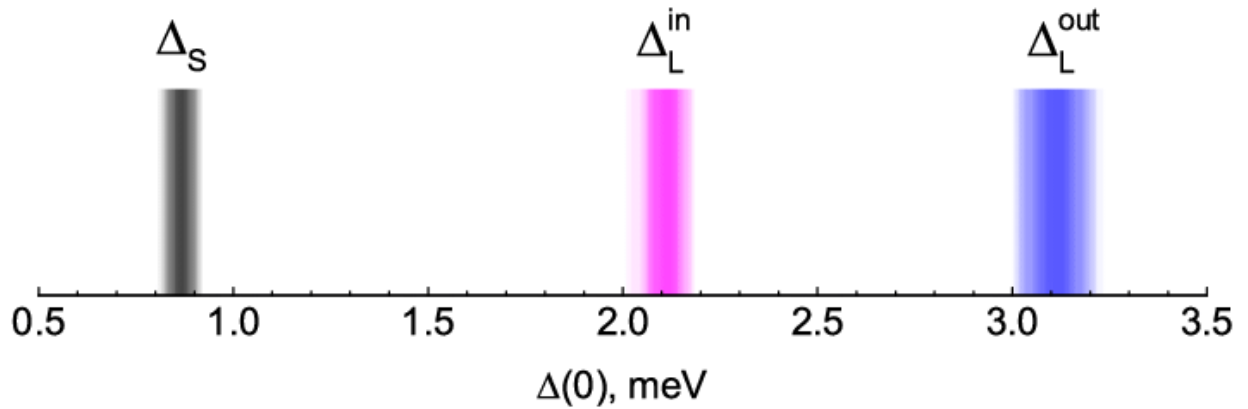

**Fig. 4.** Color histograms of the values of the superconducting order parameters at T << $T_c$ experimentally obtained in several single crystals of BaFe$_{1.86}$Ni$_{0.14}$As$_2$ from the same batch. The color intensity mapping displays the distribution of the obtained values of the energy parameters, whereas the positions of the most saturated color point to the average gap values.

At higher bias voltages |eV| ≈ 8.5-9.8 meV (orange areas, $n_{res}$=1 label) less intensive features forming a doublet are located. At half-bias |eV| \≈ 4.9 meV ($n_{res}$=2 label), a possible second subharmonic (of the outer dip of the doublet) is present, whereas the expected position |eV| ≈ 4.3 meV of that of the inner dip matches the $\Delta_L^{in}$ dip, vanishing as a distinct minimum. The origin of such fine structure is discussed below in Section D.

The complex Andreev structure detailed in Fig. 2 is *reproducible* in the dI(V)/dV spectra of Andreev arrays formed in one and the same BaFe$_{1.86}$Ni$_{0.14}$As$_2$ single crystal. Fig. 3 shows a set of dynamic conductance curves and CVC's for Andreev arrays with various number of junctions m (the second from the bottom, red curve, is the same as in Fig. 2). For each spectrum, the bias voltage values were divided by the corresponding m, and the dI(V)/dV axis is normalized by G(12.5 mV) → G(∞) at eV = 12.5 mV > 2Δ. The Andreev structures pointed to with the same notification as in Fig. 2, look similarly.

The data shown in Fig. 3 also illustrates two advantageous feasibilities provided by the used MCBJ configuration. Firstly, it is possible to vary the area of the junction (by sliding the cryogenic clefts onto the same terrace), keeping constant m. The red and cyan curves in Fig. 3 corresponding to the arrays formed under a sequent readjustment and having equal m=11 show different normal resistance per junction, as can be seen from the corresponding CVC's in Fig. 3b (thin red and bold cyan lines). On the other hand, one can readjust the number of junctions m in the array, keeping the area and transparency of the constriction almost constant. When considering the data shown by blue (m=10), cyan (m=11), and orange color (m=13), despite different $m$ numbers, the normal resistance per one junction appears the same, and the corresponding CVC's in Fig. 3b coincide.

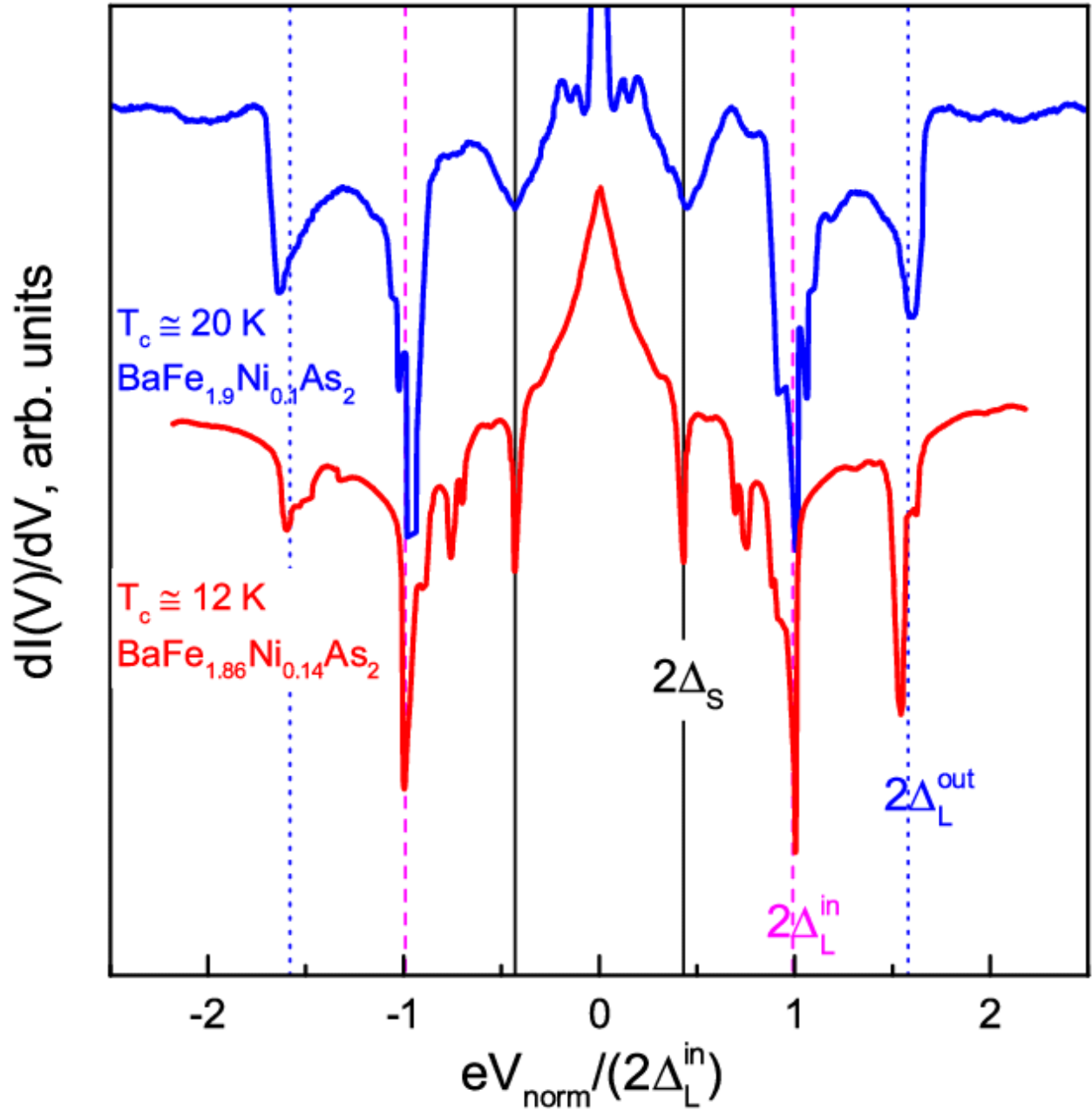

**Fig. 5.** Dynamic conductance spectrum of SnS-Andreev array obtained at T << $T_c$ in an overdoped BaFe$_{1.86}$Ni$_{0.14}$As$_2$ single crystal with $T_c^{local}$ ≈ 11.8 K (lower red curve), as compared to the blue dI(V)/dV curve obtained in optimally doped BaFe$_{1.9}$Ni$_{0.1}$As$_2$ single crystal with $T_c^{local}$ ≈ 20 K. The raw data for the Ni:0.1-sample are taken from our recent work [26]. The bias is normalized by $\Delta_L^{in}$ (being the position of the most intensive gap feature), the nonlinear $G_N$ background is suppressed for both spectra. The positions of the three fundamental harmonics correspond to the three superconducting gap parameters $2\Delta_L^{out}$ (dotted vertical lines), $2\Delta_L^{in}$ (dashed lines), and $2\Delta_S$ (solid lines).

Since the Andreev ZBC value and the number of observed Andreev dips are also similar for these spectra, the transparency of the NS boundaries seems almost equal. Anyway, the positions of the Andreev dips in the dynamic conductance spectra coincide after the bias voltage normalizing by m (see the Appendix for details), thus evidencing these features relate to the bulk properties of the superconductor and cannot be attributed to surfaces degradation, proximity effect, dimensional effects or any artifacts.

The values of three energy gap parameters at $T \ll T_c$ experimentally obtained in the probes of various $BaFe_{1.86}Ni_{0.14}As_2$ single crystals from the same batch are summarized in color histograms in Fig. 4. Each color intensity mapping displays the distribution of the obtained gap energy parameter. Clearly, in Fig. 4 the data gather into three distinct batches, where the positions of the most saturated color determine the average values of the superconducting order parameters: anisotropic large gap with $\Delta_L^{out}(0)$ = 3.11 ± 0.06 meV, $\Delta_L^{in}(0)$ = 2.10 ± 0.05 meV, and almost isotropic small gap $\Delta_S(0)$ = 0.86 ± 0.03 meV. The corresponding BCS ratios $2\Delta_L^{out}(0)/k_BT_c \approx 6$, $2\Delta_L^{in}(0)/k_BT_c \approx 4.1$ both exceed the BCS weak-coupling limit 3.5, whereas the ratio for the small gap $2\Delta_S(0)/k_BT_c \approx 6 \ll 3.5$ indicate a ``driven'' superconductivity in the bands where $\Delta_S$ develops.

The Andreev structures in the dynamic conductance spectra obtained in overdoped samples (see Figs. 2, 3) resemble those considered by us earlier in almost optimally doped samples [19,25,26]. Figure 5 shows the dI(V)/dV spectrum for Andreev array with local $T_c \approx 12$ K in $BaFe_{1.86}Ni_{0.14}As_2$ crystal (lower curve), as compared to the data with $T_c^{local} \approx 20$ K obtained in $BaFe_{1.9}Ni_{0.1}As_2$ crystal (upper curve) taken from [26]. Both spectra were measured at $T \ll T_c^{local}$. In order to collate the spectra corresponding to $T_c$ differing by almost 2 times, for each dI(V)/dV we normalize the bias voltage axis by $2\Delta_L^{in}$, being the position of the most intensive gap feature. After such normalization, it becomes clear that the positions and amplitudes of the strongest Andreev minima in the spectra (see vertical lines in Fig. 5) are similar to each other. The positions and the shapes of the fundamental harmonics of the anisotropic large gap $2\Delta_L^{out}$ and $2\Delta_L^{in}$ labels, dotted and dashed vertical lines, respectively), and isotropic small superconducting gap ($2\Delta_S$ label, solid line) are in a well agreement, thus meaning the relation between the gap magnitudes is almost constant for optimally doped and overdoped $Ba(Fe,Ni)_2As_2$. The characteristic ratios for the three gap energy parameters are also similar with those determined by us in optimally doped crystals earlier [19,25,26].

- **C. Gap temperature dependence**

Temperature evolution of the dynamic conductance spectrum shown in Fig. 2 is detailed in Fig. 6. For the curve measured at the lowest T = 2.8 K (black line) and the highest T = 14.3 K (dashed curve), the dI(V)/dV axis is normalized by $G_N(V=10mV, 2.8K}) \rightarrow G_N(\infty, 2.8K)$. As follows, the nonlinear shape of dynamic conductance background remains unchanged with temperature in the superconducting state and just above $T_c$. It indicates the general dI(V)/dV nonlinearity should not been directly associated with the superconducting phase. Other curves in Fig. 6 are shifted vertically by some constant values $c_T$ for clarity. With temperature increase, all Andreev dips in Fig. 6 become less intensive and shift toward zero, with it, ZBC peak starts to fade. The two upper curves measured at 12.4 K and 14.3 K show neither Andreev dips nor enhanced ZBC, thus corresponding to the normal state.

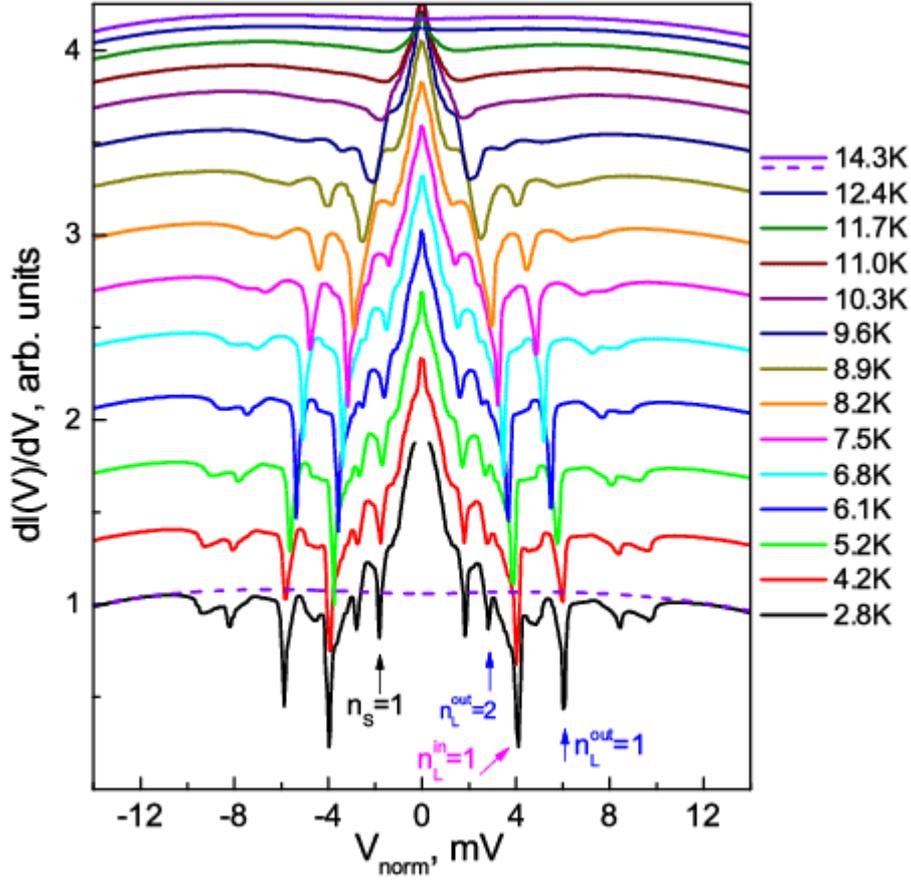

**Fig. 6.** Evolution of the dynamic conductance spectrum with temperature for SnS Andreev array from Fig. 2. For the spectra measured at 2.8 K (black line) and 14.3 K > $T_c$ (dashed line), the dynamic conductance is normalized by $G_N(V=10\ mV) \rightarrow G_N(\infty)$ at 2.8 K, thus showing the dI(V)/dV background remains unchanged with temperature up to $T_c$. Other curves are shifted vertically by some $c_T$ for clarity. The position of the gap features at 2.8 K is labelled similarly to Fig. 2. $V_{norm} = V/m$, with m=11.

Consider the position of all the superconducting state features versus temperature presented in Fig. 7. Roughly approximated T-trends of all the features seem turning to zero bias at local critical temperature $T_c^{local} \approx 11.8$ K that indicates the contact area transition to the normal state. The bias voltage of the fundamental and the second subharmonics $V(n_L^{out} =1, 2)$ of the outer edge of the large gap are shown by solid and open circles, respectively.

In accordance with Eq. 1, the doubled position of the n=2 minima has to coincide with the fundamental harmonic, since $2eV_2 = 2\Delta$. Let us check this issue. We plot such the $2V_{n=2}(T)$ dependences in Fig. 7 by connected symbols (the second column of labels). When doubling the position of the $n_L^{out}$ =2 dip (stars), it matches that of the fundamental dip (solid blue circles) at any temperature while it is observed. This means the dips labelled as $n_L^{out}$ =1, 2 indeed relate to one and the same superconducting order parameter. Moreover, such correspondence unambiguously proves the obtained temperature behavior originates from intrinsic superconducting phenomena natural to Ba(Fe,Ni)$_2$As$_2$ rather than any undesired force (for example, local excess of the critical current density would lead to the fundamental harmonic ``overheating'', i.e. $V_{n=1} < 2 V_{n=2}$). The T-behavior of the fine structure (orange symbols) is detailed in Section D below.

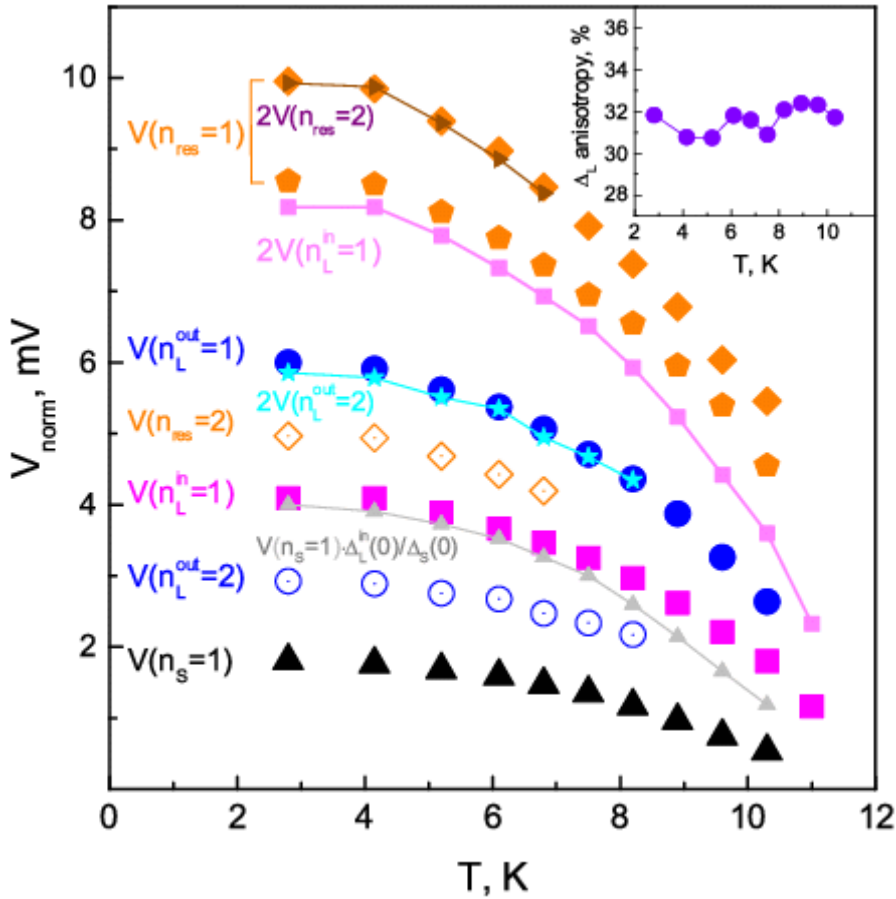

**Fig. 7.** Temperature dependence of the Andreev features in Fig. 6 versus temperature: the first (solid circles) and the second (open circles) subharmonics of $\Delta_L^{out}$ (in blue), the n=1 features of the $\Delta_L^{in}$ (magenta squares) and $\Delta_S$ (black up triangles); the positions of the high-bias doublet possibly caused by a resonant coupling with a bosonic mode (orange rhombs, pentagons), and its second subharmonic (open orange rhombs). For the n=2 Andreev features, also shown by small connected symbols are their doubled positions $2V(n_L^{out}=2)$ (cyan stars), $2V(n_{res}=2)$ (right triangles), matching the corresponding dependences of the fundamental harmonics. The $2V(n_L^{in}=1)$ dependence (small squares) and $V(n_S=1)* \Delta_L^{out}(0)/\Delta_S(0)$ (small gray triangles) are presented for comparison. The inset shows temperature dependence of the large gap anisotropy taken as $100\% * [1 - \Delta_L^{in}/\Delta_L^{out}]$.

The fundamental dip of the $\Delta_L^{in}$ SGS shows temperature behaviour (squares in Fig. 8) similar to the outer edge of the large gap. The inset exhibits the temperature dependence of the anisotropy defined as $100\% * [1 - \Delta_L^{in}/\Delta_L^{out}]$. With temperature increase almost until $T_c$, the large gap anisotropy is nearly constant, keeping within the range 31%-33%. Similar temperature independence of the $\Delta_L$ anisotropy was reproducibly observed by us earlier in optimally doped crystals [19,25,26]. Additionally, it excludes an attributing of the $n_L^{in} = 1$ dips (squares) to interference $\Delta_L^{out} + )/\Delta_S$ feature, or to in-gap Andreev bound states manifestation due to the very different temperature dependences of corresponding minima in both cases.

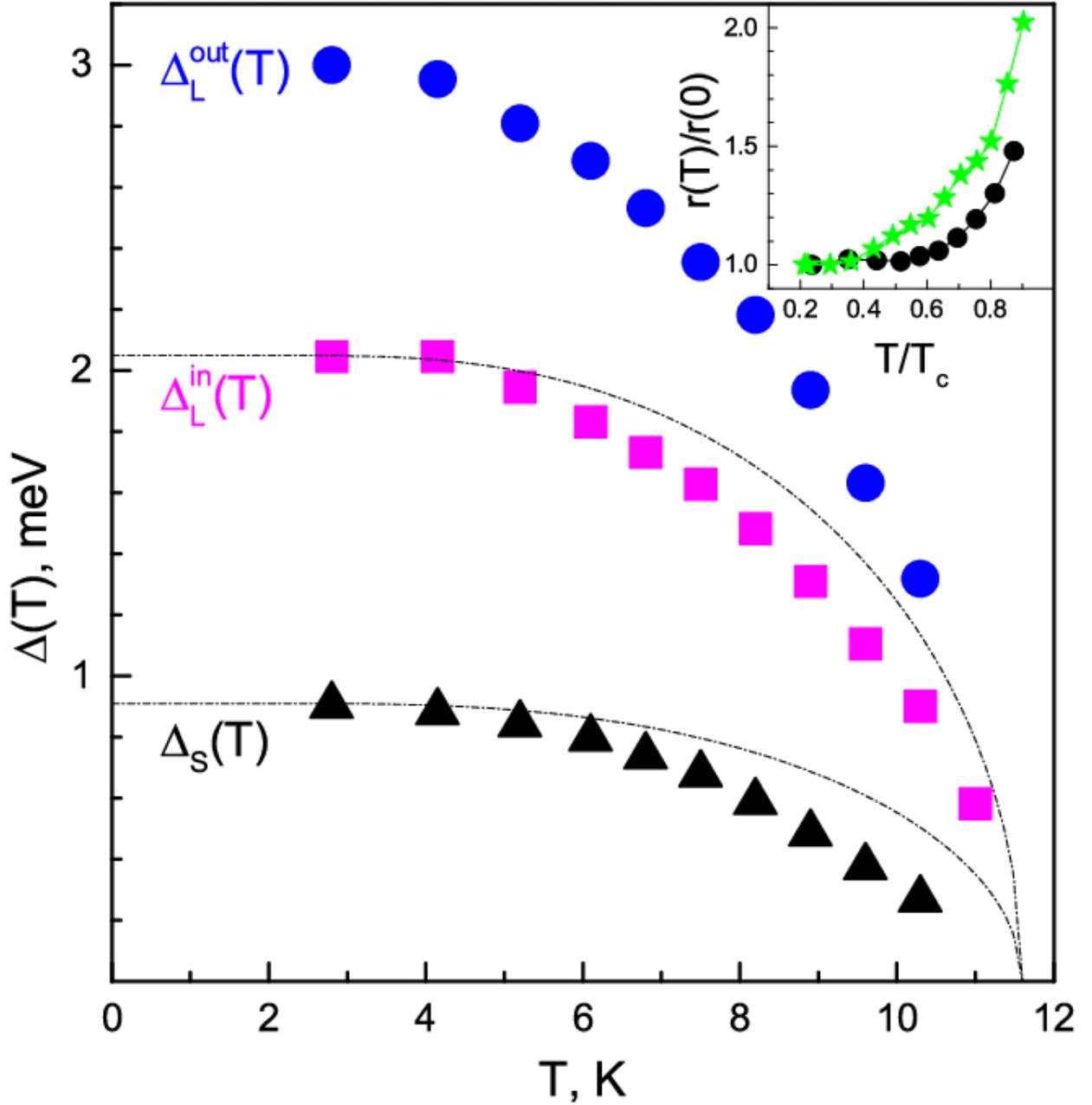

**Fig. 8.** Temperature dependence of the superconducting order parameters $\Delta_L^{out}$ (circles), $\Delta_L^{in}$ (squares), and $\Delta_S$ (triangles) obtained using data in Fig. 7. The single-band BCS-like curves (dash-dot line) are shown for comparison. In the inset, the gaps ratio $r(T)/r(0) = [\Delta_L^{in}(T)/\Delta_L^{in}(0)] / [\Delta_S(T)/\Delta_S(0)]$ is shown versus normalized temperature for this contact (black circles) and for an SnS-contact formed in optimally doped Ni:0.1 compound (green stars).

The position of the $n_S=1$ dip of the small gap (black up triangles in Fig. 1) decreases a bit faster in the vicinity of $T_c$, as becomes clear when superposing it with the position of the $n_L^{in} = 1$ dip by the normalization to its value at the lowest T (the result is shown by small up triangles connected by the gray line).

Temperature evolution of the superconducting gap parameters obtained using Fig. 7 is presented in Fig. 8. The dependences of the superconducting gaps deviate from single-band BCS-like curves (dash-dot lines), passing below them. Accounting the shapes of the $\Delta_{L,S}(T)$ curves are rather similar, one may assume a strong interband coupling being one order of magnitude as the intraband one, i.e. $\beta = \lambda_{LL} \lambda_{SS} / (\lambda_{LS} \lambda_{SL}) \approx 1$, where $\lambda_{ij}$ are the coupling constants. Note in case of zero determinant of the $\lambda_{ij}$ square matrix $\det(\lambda_{ij}) = 0$ and therefore $\beta = 1$, the ratio $\Delta_i(T)/\Delta_j(T)$ is constant within the whole temperature range up to $T_c$ [79,80].

Figure 8 also resembles the temperature dependences of the gap energy parameters obtained by us earlier in $BaFe_{1.9}Ni_{0.1}As_2$ single crystals with almost optimal composition and $T_c \to 18\text{-}20$ K [19,25,26]. Nonetheless, in Ni:0.1 samples, the small gap tends to $T_c$ noteworthy faster, showing almost linear decreasing with temperature (see Fig. 7 in [25] and Figs. 3 in [19,26]). In other words, the ratio between the large and the small gaps $r(T)/r(0) = [\Delta_L^{in}(T)/\Delta_L^{in}(0)] / [\Delta_S(T)/\Delta_S(0)]$ rapidly increases toward $T_c$ (stars in the inset of Fig. 8) thus unambiguously identifying $\Delta_S$ as a distinct superconducting order parameter. In overdoped Ni:0.14 samples, such ratio $r(T)/r(0)$ demonstrates a bit weaker temperature dependence (black circles in the inset). Nonetheless, as shown in Fig. 5, at $T \ll T_c$ the $\Delta_L^{in}(0)/\Delta_S(0)$ ratio is almost the same, and the $dI(V)/dV$ minima in the red lower curve marked by solid lines can be considered as having the same nature $V(n_S=1)$ as those in the blue curve. Such argument additionally favors an existence of the distinct small superconducting gap $\Delta_S$ in overdoped samples. Therefore, the observed here similarity between $\Delta_L(T)$ and $\Delta_S(T)$ dependences in Ni:0.14 compound seems indicating relatively strong ($\beta \to 1$) interband coupling in overdoped $Ba(Fe,Ni)_2As_2$ and its increase in comparison with optimally doped Ba-122 ($\beta \gg 1$) estimated in [25]).

- **D. The overgap fine structure**

In general, an Andreev carrier could lose its energy by emitting a boson during MAR process, in case the boson energy $\varepsilon_0 \leq 2\Delta$. As a result of a single boson emission, the $dI(V)/dV$ spectrum would show satellite dips beyond the parent gap subharmonics at positions

$$|eV_n^{res}| = (2\Delta + \varepsilon_0)/n \qquad (3)$$

in case of isotropic gap [68], whereas for n=1 the boson energy is the ``distance'' between the fine structure feature and the fundamental 2Δ-feature. Such fine Andreev structure was earlier observed by us in magnesium diborides [69,70] and attributed to Leggett plasmon [71] emission. In iron-based superconductors, Leggett plasma modes are believed to be unobservable since having too high energy [72,73]. Accounting the widely observed spin resonance in various Ba-122 compounds, one may suppose, for example, an emission of a spin exciton. According to theoretical calculations in the framework of the $s^{\pm}$ approach [74,75], the energy of the spin exciton $\varepsilon_0$ roughly corresponding to the position of the spin resonance, at $T \to 0$ does not exceed the indirect gap between the nested Fermi surfaces, and shifts toward zero in case of anisotropic gap [75]. In various oxypnictides of the 1111 family studied by us, the $dI(V)/dV$ spectra of Andreev contacts showed the SGS accompanied by a fine structure, whereas the extracted boson energy did not exceed $\Delta_L(0) + \Delta_S(0)$ and scaled with $T_c$ with the characteristic ratio $\varepsilon_0/k_BT_c \approx 3.3$ [50,76,77].

For overdoped BaFe$_{1.86}$Ni$_{0.14}$As$_2$, temperature behavior of the fine structure doublet located above the large gap is shown in Fig. 7 by orange rhombs (the outer dip), pentagons (the inner dip), and open rhombs (the second $n_{res}= 2$ subharmonic of the outer dip). The doubled position of the $n_{res}= 2$ feature (connected right triangles) matches exactly the dependence of its parent $n_{res}= 1$ harmonic, thus in accordance with Eq. 1 these features are attributed to one and the same SGS. Although temperature dependence of the fine structure resembles those of the superconducting gaps, the $\Delta_{L,S}(T)$ evolve with T a bit weaker. As mentioned above, at T << $T_c$ the positions of the inner dip of the doublet and the minimum related to $2\Delta_L^{in}$ differ by roughly two times, and hence it may seem they form an SGS as n=1 and n=2.

Turning to the temperature evolution, it becomes clear that the doubled position of the $\Delta_L^{in}$ fundamental harmonic (small connected squares) more and more deviates from the $n_{res}= 1$ dip toward $T_c$. Judging by these reasonings, (a) we do not directly relate the fine structure to the superconducting order parameters, but attribute it to some resonant effect, for example, a coupling with a bosonic mode developing in the superconducting state only.

(b) We argue, the fine structure cannot be caused by any geometric resonance since its position and the shape is almost independent on the normal resistance (thus the contact area) and the (random) number of junctions in the array, see Fig. 3. (c) As well, phonon emission (the corresponding estimation of the phonon energy gives 20, 32 cm$^{-1}$) also seems hardly possible: we have never observed such resonant features in SnS contacts formed in various superconductors (in particular, in high-temperature cuprates, the low-transparency Josephson (break)junctions reproducibly showed phonon resonances, those fully absent in high-transparent SnS junctions [51-53]). (d) The backscattering effect discussed in Section A also could not be the cause, accounting the significant amplitude of the fine structure (up to $0.2G_N$) and its observability below $T_c$ only (on the contrary, the backscattering effect is not directly related to the superconducting state). (e) It is impossible to interpret the fine structure as any possible in-gap states (Andreev bound states typical for coherent transport regime, Yu-Shiba-Rusinov states, etc.): since their energy $|\varepsilon^{BS}|$ is less than $\Delta$, the corresponding subharmonic fine structure would be located at $|\varepsilon_n^{BS}| = 2\varepsilon^{BS}/n$. Obviously, for the n=1 feature $|eV_1^{BS}| < 2\Delta$, on the contrary, the experimentally observed fine structure is overgap.

In Ba-122, a spin exciton emission during MAR seems rather feasible, although due to a possible gap anisotropy and the doublet-shaped fine structure, the extraction of the boson energy and its temperature dependence is ambiguous. Seemingly, the observed here in overdoped Ba-122 doublet dI(V)/dV fine structure feature cannot originate from the doublet spin resonance reported in underdoped samples barely [16,78], and thus needs in another explanation. Since the distance between the $2\Delta_L^{in}$ feature and the bosonic feature appears larger than $2\Delta_L^{in}(0)$ (see blue vertical bar in Fig. 3, $n_L^{out}=1$ label), the boson emission is possible for the momentum directions corresponding to $\Delta_L^{out}$ only. Therefore, the energy of the bosonic mode at T << $T_c$ could be estimated as the difference between the position of the bosonic doublet and $2\Delta_L^{out}(0)$ in accordance with Eq. 3. Taking the edge positions of the bosonic doublet, we get the two feasible values of the boson energy $\varepsilon_0' \approx 2.5$ meV and $\varepsilon_0'' \approx 4$ meV at T << $T_c$. If supposing the $\varepsilon_0(k)$ dependence in the momentum space, such anisotropy of the boson energy could possibly give the energy range $\varepsilon_0(k) \approx 2.5 - 4$ meV. In any case, the estimated boson energy $\varepsilon_0 \leq \Delta_L^{out}(0) + \Delta_S(0)$ at $T \ll T_c$ agrees with the theoretical predictions for a spin exciton mode [74,75].

As well, any resonant electron-boson interaction could renormalize electron DOS with an appearance of DOS features outside the gap. If so, the DOS peaks could also be a reason of the observed fine structure. In such case, additional theoretical studies are necessarily in order to determine the boson energy from the experimental data.

**Conclusions**

By using multiple Andreev reflection (MAR) spectroscopy of classical SnS contacts (with incoherent transport) produced by the mechanically controlled break-junction (MCBJ) technique, we directly probed the structure of the bulk superconducting order parameter in overdoped BaFe$_{1.86}$Ni$_{0.14}$As$_2$ single crystals with critical temperature $T_c \approx 12$ K. At $T \ll T_c$, we revealed anisotropic but nodeless large gap $\Delta_L^{out}$ = 3.11 ± 0.06 meV, $\Delta_L^{in}$ = 2.10 ± 0.05 meV with the characteristic ratios about 6 and 4, respectively, and isotropic small gap $\Delta_S$ = 0.86 ± 0.0 meV with $2\Delta_S/k_BT_c \approx 2$. The temperature dependence of the superconducting gaps favors rather strong interband coupling, comparable to the interband one ($\lambda_{LL} \lambda_{SS} \approx \lambda_{LS} \lambda_{SL}$). The observed overgap fine structure could result from DOS features or from a resonant emission of bosons with two distinct energies $\varepsilon_0' \approx 2.5$ meV and $\varepsilon_0'' \approx 4.0$ meV or with momentum-dependent energy $\varepsilon_0(k) \approx 2.5 - 4.0$ meV. The nonlinear features of the background in the normal and the superconducting state could be considered as fingerprints of unconventional DOS energy distribution in the vicinity of the Fermi level.

**Acknowledgments**

TEK acknowledges the support from RSF (project no. 19-72-00196). SAK, KSP and VAV were supported by the state assignment of the Ministry of Science and Higher Education of the Russian Federation (topic ``Physics of high-temperature superconductors and novel quantum materials'', no. 0023-2019-0005). The research has been partly done using the research equipment of the Shared facility center at Lebedev Physical Institute RAS.

# Appendix: The Details of Extracting the Energy Parameters Using Andreev Spectroscopy of MCBJ's

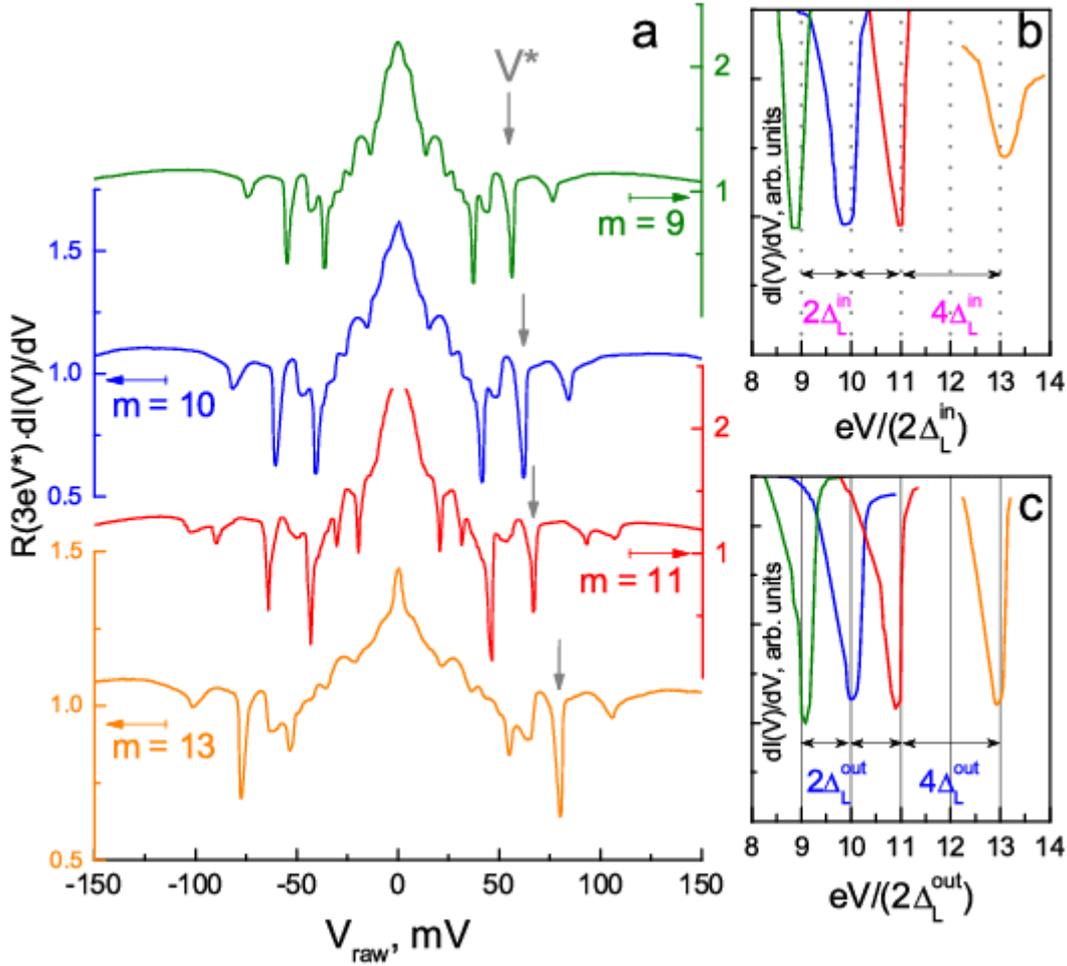

**Fig. 9.** a) A set of raw dynamic conductance spectra for SnS-Andreev arrays with various number of junctions m. Gray arrows point to the position of the fundamental harmonic of the largest gap $\Delta_L^{out}$. For each spectrum, the dI(V)/dV is normalized by that at V = 3V*. (b,c) The fragments of the spectra from (a) containing Andreev dips of the large gap edges. The fragments are shifted vertically for comparison. In (b), the bias voltage axis is divided by $\Delta_L^{in}$, whereas vertical dashed lines indicate the positions eV=m * $2\Delta_L^{in}$. In (c), similar was done for the Andreev dips of the $\Delta_L^{out}$ (labelled as V* in (a)).

When cracking a sample with a layered crystal structure, arrays are naturally formed, possibly developed on the steps-and-terraces of the cryogenic cleft. Due to the possibility of probing the *bulk* properties, spectroscopy of such stacks seems rather advantageous, as compared with single junction study; although, one more intermediate purpose raises, namely to determine the number of junctions in each formed array. Primarily, in order to solve this problem, a large data statistics is essential. As mentioned above, the raw position of the Andreev feature becomes scaled by a factor of natural but accidental m, hence, the raw dI(V)/dV spectrum provides 2Δ*m energy value. The details of the Δ value extraction from the raw experimental data are presented in Figs. 9, 10.

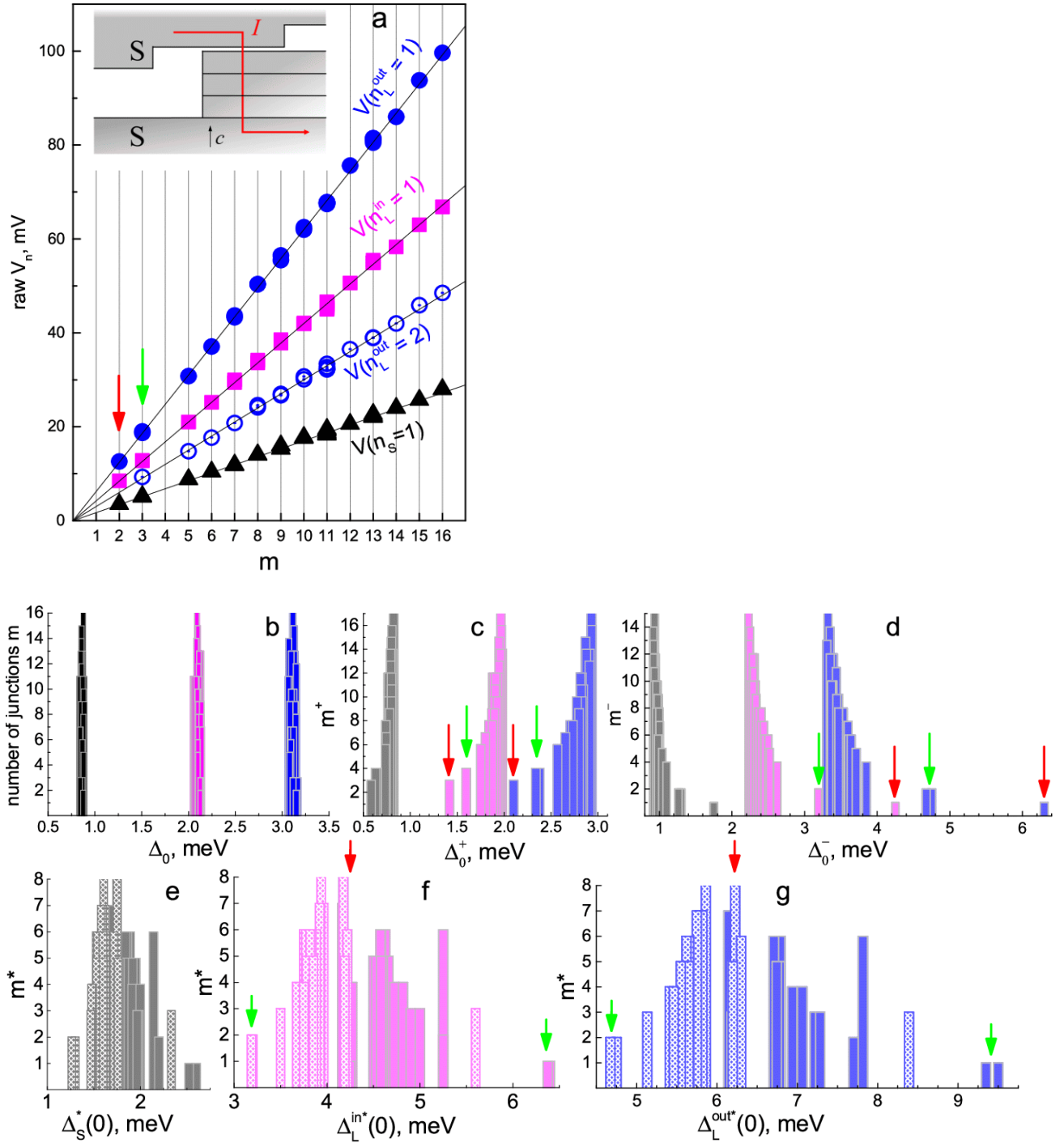

**Fig. 10.** a) The raw positions of the Andreev subharmonics at T << $T_c$ of the three superconducting order parameters in the obtained dI(V)/dV spectra versus the congruent m number of junctions in the stack. The extracted $\Delta_L^{out}$ = 3.11 ± 0.06 meV, $\Delta_L^{in}$ = 2.10 ± 0.05 meV, $\Delta_S$ = 0.86 ± 0.03 meV, values are shown in (b). A presumable sketch of ScSc-…-S array (m=4) developed in the layered single crystal is shown in the inset of (a). The other possibilities to normalize $V_n$ values are presented in panels (c-g). For comparison, the large gap values to be obtained when the raw $V_n$ are normalized with shifted sets $m^+$ = m+1, and $m^-$ = m-1 (c,d); using the halved set m*=m/2, the following values rounded down (solid bars) and up (dashed bars) are shown in (e-g) panels. Each bar height in (b-g) corresponds to the chosen m. Arrows point to the data obtained for m=2, 3 junction stacks.

A representative set of raw dynamic conductance spectra of Andreev arrays is shown in Fig. 9. The blue, red, and orange curves correspond to those shown in Fig. 3, whereas the top curve in Fig. 9 relates to an array with m=9 junctions. The raw position V* of the fundamental harmonic of the largest gap parameter $2\Delta_L^{out}$ is pointed by arrows. For each spectrum, the absolute dynamic conductance is normalized by that at eV = 3V* = $6\Delta_L^{out}$ Under fine mechanical readjustment, the touching point between two cryogenic clefts often switches to neighbour terraces, thus resulting in the m change by 1-2. If such happens, the position of the fundamental Andreev dip n=1 correspondingly shifts by $2\Delta$ or $4\Delta$, thus facilitating the determination of the gap energy quantum. Fig. 9b shows the fragments of the spectra from (a) panel containing the intensive Andreev feature corresponding to $2\Delta_L^{in}$. Clearly, the position of these dips in (b) is quantized as illustrated by dotted lines. Taking this quant as $2\Delta$, one makes sure that each position of the Andreev dip is also a multiple of it: as normalizing the bias voltage axis by the determined $2\Delta_L^{in}$ the dips line up related to the natural numbers m. In (c), similar was done for the Andreev dips pointed to as V* in (a) and corresponding to the largest gap parameter $\Delta_L^{out}$. For all the curves, similar scaling is observed, thus resulting in the same set of m. Note if supposing a twice smaller $2\Delta_L^{out}$ quant (which would not break the quantization but would result in doubling of the determined m numbers), one gets the characteristic ratio $2\Delta_L^{out}/k_BT_c < 3.5$ which is impossible for the largest order parameter in a multiple-gap superconductor.

For the Andreev spectra of the arrays formed in the x=0.14 single crystals from the same batch, the positions of the main gap features were arranged in ascending order versus the assigned natural numbers m, as shown in Fig. 10a. All the dependences are well fitted with straight lines crossing the origin. The slope of the curve showing the n=2 Andreev dip position of the $\Delta_L^{out}$ appears twice lower than that of the fundamental n=1 dip, in accordance with Eq. 1. The multiplying of the Andreev feature positions with m indicates that all the determined gap parameters have a bulk nature (on the contrary, those of a surface gap would not scale with m).

The statistics on gap values at T << $T_c$ (i.e. the raw position of the fundamental harmonics shown in (a) divided by 2 and by the selected m) is summarized in Fig. 10b. Each triple of the equal-height bars represents a certain array, with the bar height corresponding to the number of junctions in accordance with (a). The extracted order parameters are $\Delta_L^{out}$ = 3.11 ± 0.06 meV, $\Delta_L^{in}$ = 2.10 ± 0.05 meV, $\Delta_S$ = 0.86 ± 0.03 meV. Remarkably, the selected set of m provides the $\Delta_0$ value uncertainty about 2%-3%, thus proving the selected m set is correct.

Noteworthily, there is no correlation between the gap value and the corresponding m, which also proves the energy parameters are insensitive to the contact configuration and characterize the fundamental properties of the superconductor.

For comparison, the gap values to be obtained when using another sets of m are shown in Fig. 10(c-g). Panels (c,d) expose the distributions to be obtained if assuming shifted $m^+$ = m+1 and $m^-$ = m-1 sets, respectively. These assumptions scatter the normalized value of the energy gaps. This becomes obvious especially for the arrays with the lowest m=2, 3 estimated (see arrows in panels (c) and (d)). Panels (e-g) depict the gap distributions of the normalized superconducting gaps $\Delta_0^*$ to be obtained if assuming the m=2 junctions to be in reality a single junction with m* =m/2 rounded down (solid bars) and rounded up (dashed bars) alternative sets. Under such normalizing, the contacts with even m still provide reproducible and non-scattered but double $\Delta_0$* values, whereas the odd m numbers are to be rounded, thus providing ambiguous normalized value of the SC gaps especially for the lowest m=3 estimated (see green arrows in panels (f) and (g)). Anyhow, all the cases supposed in (c-g), clearly, provide strongly scattered gap values, with obvious tendency $\Delta^{+/-/*}\to$ true $\Delta_0$ with $m^{+/-/*}$ increase, thus supposing badly wrong $\Delta_0$ correlation with the array properties. Thereby, the m set derived from Figs. 9 and 10a was used to normalize the dI(V)/dV spectra shown in Figs. 2, 3.